\documentclass[10pt, conference]{IEEEtran}
\IEEEoverridecommandlockouts
\usepackage{graphicx} 

\usepackage{amsmath}
\usepackage{xcolor}
\usepackage{comment}

\title{High-Performance Pipelined NTT Accelerators with Homogeneous Digit-Serial Modulo Arithmetic}
\author{%
\IEEEauthorblockN{George Alexakis}
\IEEEauthorblockA{Electrical and Computer Engineering\\ 
Democritus University of Thrace, Xanthi, Greece}
\and
\IEEEauthorblockN{Dimitrios Schoinianakis}
\IEEEauthorblockA{Nokia Bell Labs\\  
Athens, Greece}
\and
\IEEEauthorblockN{Giorgos Dimitrakopoulos}
\IEEEauthorblockA{Electrical and Computer Engineering\\ 
Democritus University of Thrace, Xanthi, Greece}}

\begin{document}

\maketitle

\begin{abstract}
The Number Theoretic Transform (NTT) is a fundamental operation in privacy-preserving technologies, particularly within fully homomorphic encryption (FHE). The efficiency of NTT computation directly impacts the overall performance of FHE, making hardware acceleration a critical 
technology that will enable realistic FHE applications. Custom accelerators, in FPGAs or ASICs, offer significant performance advantages due to their ability to exploit massive parallelism and specialized optimizations. However, the operation of NTT over large moduli requires 
large word-length modulo arithmetic that limits achievable clock frequencies in hardware and increases hardware area costs. To overcome such deficits, digit-serial arithmetic has been 
explored for modular multiplication and addition independently. The goal of this work is to leverage digit-serial modulo arithmetic combined with appropriate redundant data representation to design modular pipelined NTT accelerators that operate uniformly on arbitrary small digits, without the need for intermediate (de)serialization. The proposed architecture enables high clock frequencies through regular pipelining while maintaining parallelism. Experimental results demonstrate that the proposed approach outperforms state-of-the-art implementations and reduces  hardware complexity under equal performance and input-output bandwidth constraints.
\end{abstract}

\section{Introduction}
\label{s:intro}
Privacy-preserving technologies (PPT) and post-quantum cryptography (PQC) are two of the most anticipated technologies that promise to revolutionize data monetization and secure our communications in the long-term, especially in the realm of the 6G era, which is expected by the end of this decade \cite{nokia_sec_6G}. 6G will support PQC from day one, whereas advanced PPTs like Fully Homomorphic Encryption (FHE) will offer the means to unlock new data monetization opportunities, advanced machine learning (ML), and AI use-cases \cite{nokia_sec_6G}.

At the core of these applications lies the Number Theoretic Transform (NTT)~\cite{liang2022number}. 
Especially for FHE, accelerating NTT is crucial in order to achieve an acceptable performance, considering that current projections refer to a $\times 5000$ speed-up in performance if pragmatic FHE use-cases are to be developed \cite{f1}.
This is primarily due to the ability of NTT to handle polynomial arithmetic efficiently, which is essential for the operations involved in these cryptographic schemes. In FHE for example, computations on encrypted data are performed by operating on polynomials that represent encrypted values. The NTT optimizes these polynomial operations by enabling faster multiplication and modular reduction. Given the high computational demands of FHE, which often involve large-scale polynomial operations, the NTT offers a much-needed efficiency boost, allowing for complex computations on encrypted data.

To achieve this goal, accelerating NTT using custom hardware accelerators, such as FPGAs or ASICs, is a viable choice due to the massive parallelism and specialized optimizations available in dedicated hardware. Custom accelerators can optimize memory access patterns, minimize data movement overhead, and implement highly efficient modular arithmetic, leading to orders-of-magnitude speedups in execution time and energy efficiency. 

The large word-length of operands found in state-of-the-art NTT applications \cite{cheon2017homomorphic} directly impacts hardware performance by increasing the complexity of modular arithmetic operations that must handle high-precision computations, requiring multi-cycle processing or wide datapaths. This increased operand size leads to longer combinational delay in arithmetic circuits, particularly in modular reduction, found in Barrett or Montgomery reduction techniques \cite{montgomery}. As a result, critical paths on hardware grow longer, limiting the maximum achievable clock frequency. Furthermore, large word-length operations demand extensive routing and storage resources, increasing power consumption and area costs and decreasing the overall modularity of the physical layout. These constraints hinder scalability and efficiency, especially in parallelized architectures, making it challenging to achieve the low-latency execution requirements of cryptographic applications.

Residue Number System (RNS) and digit-serial modular arithmetic address the problem of large word-length operands in NTT by breaking computations into smaller computing channels. RNS decomposes large integers into residues modulo smaller co-prime bases, allowing operations to be performed in parallel on smaller word sizes, effectively eliminating costly large-bit-width multiplications and modular reductions \cite{5219515, cheon2018full}. This reduces critical path delays and enhances scalability. Digit-serial arithmetic, on the other hand, processes operands in small digit-sized chunks over multiple cycles instead of handling full-width numbers at once \cite{4176855}. This approach reduces combinational logic depth, enabling higher clock frequencies while maintaining throughput. Both techniques improve performance, reduce hardware complexity, and optimize resource usage in NTT accelerators.

In this work, we focus on digit-serial arithmetic as a more scalable approach to computing NTTs using arbitrarily small bit-width words. Our goal is to leverage existing digit-serial modular arithmetic units \cite{4176855}, which have been independently designed for multiplication and addition, along with newly introduced redundant data representations, to develop modular pipelined NTT hardware architectures. These architectures operate on uniform digit sizes without requiring (de)serialization or large word-length modular reductions at any part of the NTT pipeline. The contributions of this work are summarized as follows:
\begin{itemize}
\item Pipelined NTT accelerators that operate with a uniform digit size are proposed, eliminating the need for intermediate modulo reductions. This is achieved through a carefully chosen redundant data representation that uniformly handles input, output, and all intermediate results. 
\item The introduction of a uniform digit size enables a novel and seamless integration of systolic digit-serial Montgomery multipliers into pipelined NTT accelerators, resulting in high-speed implementations.
\item The proposed NTT accelerators were implemented as open source designs using the OpenRoad physical design flow, based on the ASAP 7 nm technology library. Results from physical synthesis demonstrate the scalability of the proposed approach. The use of uniform digit sizes and systolic Montgomery multipliers enables higher clock frequencies while reducing power without increasing execution time. Compared to recent state-of-the-art NTT accelerators, the maximum clock frequency is increased by 33\%, while power consumption is reduced by 30\% on average for various NTT sizes.
\end{itemize}

\section{NTT Computation and State-of-the-art Hardware Accelerators}
\label{s:background}

The NTT transforms a sequence of $n$ coefficients of a polynomial into $n$ evaluations of that polynomial at specific points within the finite field. Therefore, multiplying two polynomials in this ``evaluation domain'' transforms the convolution-like operation of direct polynomial multiplication to an element-wise multiplication \cite{cohen}.
An inverse NTT (INTT) then transforms the result back to the coefficient domain, yielding the coefficients of the product polynomial. 

Given a sequence of $n$ integers $a = (a_0, a_1, ..., a_{n-1})$ and a primitive $n$-th root of unity $\omega$ modulo a prime $Q$ (meaning $\omega^n \equiv 1 \pmod{Q}$ and $\omega^k \not\equiv 1 \pmod{Q}$ for $1 \le k < n$), the forward NTT $A = (A_0, A_1, ..., A_{n-1})$ is defined as:
\begin{equation*}
A_k = \sum_{j=0}^{n-1} a_j \omega^{jk} \pmod{Q} \quad \text{for } k = 0, 1, ..., n-1
\end{equation*}
The INTT recovers the original sequence:
\begin{equation*}
a_j = n^{-1} \sum_{k=0}^{n-1} A_k \omega^{-jk} \pmod{Q} \quad \text{for } j = 0, 1, ..., n-1
\end{equation*}
Here, $n^{-1}$ is the modular multiplicative inverse of $n$ modulo $Q$, i.e., $n\cdot n^{-1} \equiv 1 \mod Q$.
Inputs $a_i$ are $W$-bits wide where $W = \lceil\log_2(Q)\rceil$.

The direct computation of the NTT and INTT has quadratic complexity. However, similar to the Fast Fourier Transform (FFT), the NTT can be computed with logarithmic complexity using a divide-and-conquer approach~\cite{kaya2023post}. In this case, NTT of size $n$ is recursively divided into two NTTs of size $n/2$. The results are then combined using ``twiddle factors'', which are powers of the primitive $n$-th root of unity $\omega$. This recursive divide-in-half computation pattern concludes in a logarithmic number of steps and is computed using the well known FFT butterfly structures~\cite{heckbert1995fourier}. 

\begin{figure}[t]
\centering
\includegraphics[width=0.8\columnwidth]{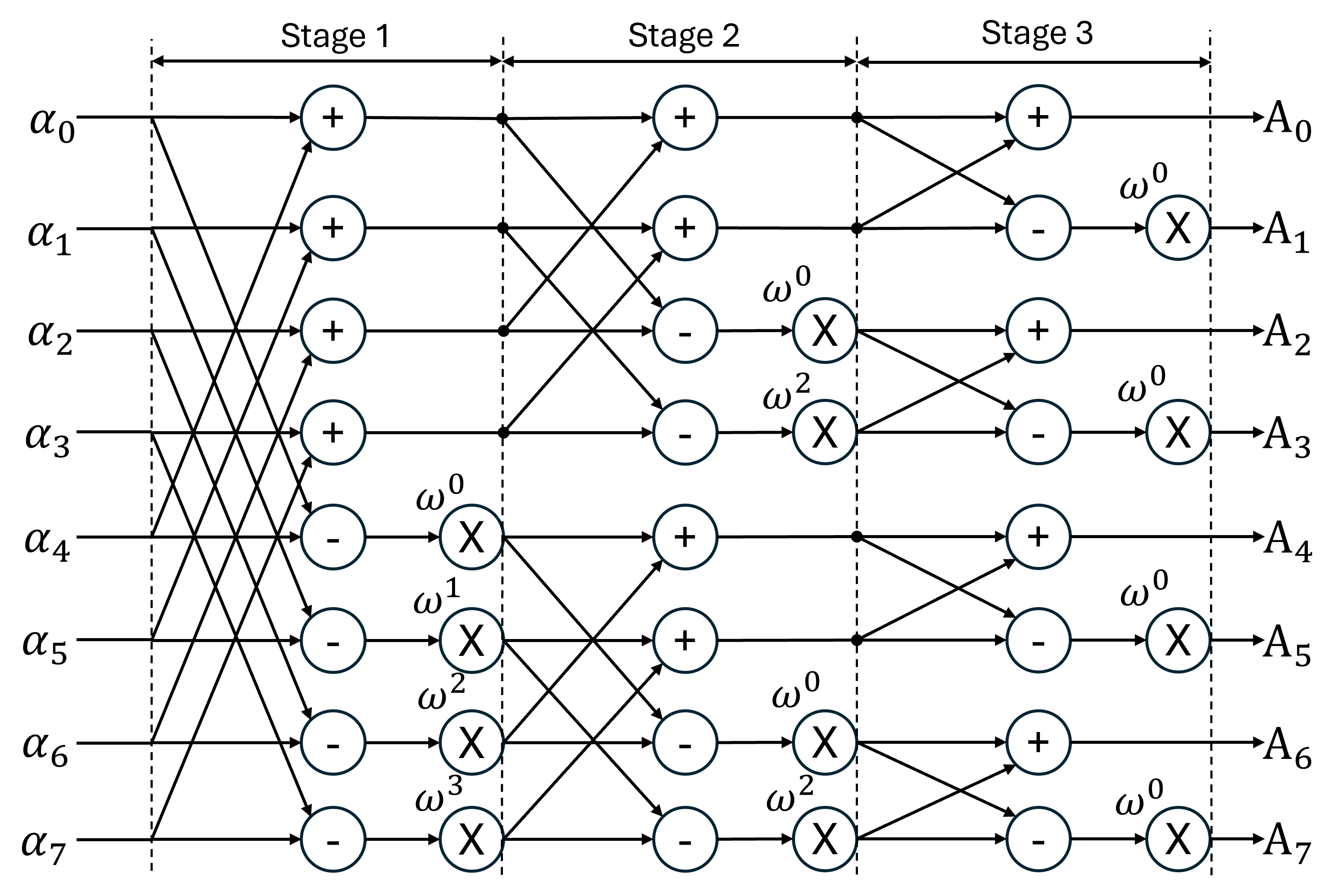}
\caption{Fast NTT computation following the Fast-Fourier Transform recursive paradigm.}
\label{fig:nttflow}
\end{figure}

Consider two adjacent elements in an intermediate stage of the NTT computation, say $A_j$ and $A_{j+n/2}$ (where $n$ is the current size of the sub-transform). A radix-2 butterfly operation computes the next stage values $A'_j$ and $A'_{j+n/2}$ as follows:
\begin{equation*}
\begin{aligned}
A'_j &= A_j + \omega^k \cdot A_{j+n/2} \pmod{Q} \\
A'_{j+n/2} &= A_j - \omega^k \cdot A_{j+n/2} \pmod{Q}
\end{aligned}
\end{equation*}
Here, $\omega^k$ is the twiddle factor, where the exponent $k$ depends on the stage of the algorithm and the index $j$. 
The twiddle factors used in each stage are specific powers of the primitive $n$-th root of unity, carefully chosen to ensure the correctness of the transform.
The entire NTT Dataflow for an 8-input using butterfly operations is shown in Fig.~\ref{fig:nttflow}. 

In essence, the NTT butterfly structure is the basic arithmetic unit that efficiently combines the results of smaller NTT computations, driven by the properties of the primitive roots of unity in the finite field.
While radix-2 is the most common, other radices (like radix-4, radix-8) can also be used in NTT algorithms. These higher-radix butterflies process more than two inputs and outputs at a time, thus reducing the number of stages and possibly improving performance.

\subsection{Parallel NTT accelerators}
NTT accelerators are designed using various approaches. A generic method employs an array of butterfly units and multi-banked memories, supported by appropriate address generation logic, enabling the computation of NTTs of arbitrary sizes in a programmable manner.
This approach is followed by~\cite{9882331} where 
modular multiplication is executed using Montgomery’s method. CoHA-NTT \cite{derya2022coha} follows a similar structure but optimizes the modular multiplication units by implementing Montgomery reduction separately. After the initial multiplication, the output is forwarded to a reduction unit consisting of several partial reduction units, each responsible for reducing its input word by a specific number of bits until the result falls within the desired range.

Although these architectures are generic, they suffer from irregular memory accesses and data transfers due to the structure of the NTT. To address this, complex reordering schemes~\cite{10843144} are implemented to ensure that elements that need to be fetched in parallel are not stored in the same memory bank. Since butterfly pairings change with each iteration, this reordering must be performed repeatedly throughout the NTT operation. A hierarchical approach was followed in~\cite{11010124} to simplify the memory accesses between NTT stages.

\subsection{Pipelined NTT Hardware Accelerators}
In contrast to generic parallel architectures, pipelined architectures follow a linear pipeline structure that targets the removal of irregular memory accesses by appropriately delaying data elements (inputs or intermediate results) before being used in the following stages of the NTT similar to high-performance FFT accelerators~\cite{garrido2022survey}.

\begin{figure}[t!]
\centering
\includegraphics[width=\linewidth]{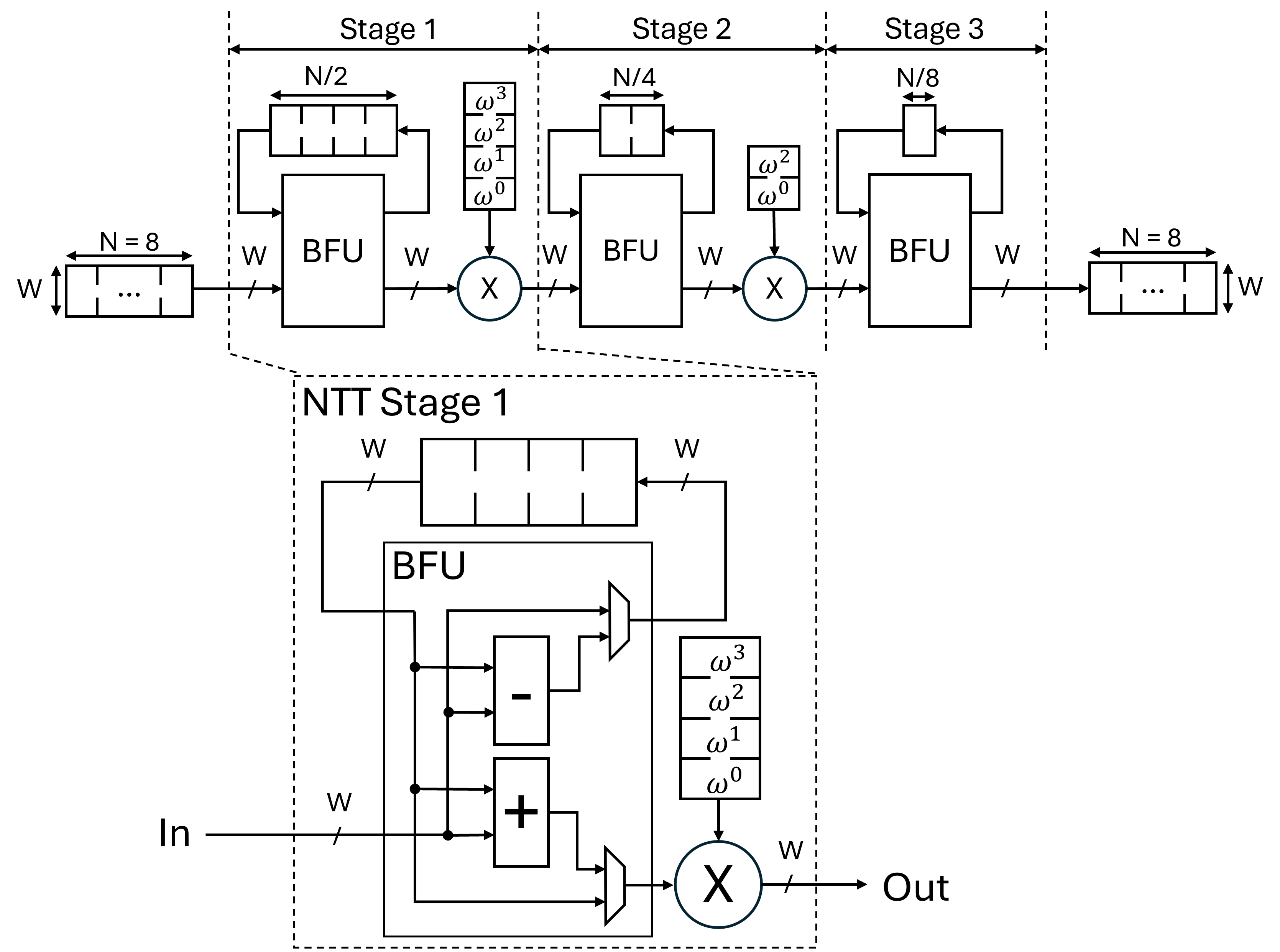}
\caption{Pipelined architecture for $N=8$ point NTT resembling the Single-path Delay Feedback architecture proposed in~\cite{garrido2022survey} for FFT.}
\label{fig:sdf}
\end{figure}

The most characteristic example of pipelined NTT architectures is shown in Fig.~\ref{fig:sdf}. The NTT accelerator consists of a logarithmic number of stages, where each one performs the necessary butterfly operation and buffers all intermediate results in order to align their transfer to the next stage with the total order of computation. 
Each NTT stage alternates between two different states at constant intervals. The first is called the move state, during which elements are simply moved in and out of the buffer of the NTT stage. The second is the compute state, where the adder and subtractor compute new results that are either stored in the buffer or forwarded to the multiplier.

Proteus~\cite{hirner2024proteus} follows this pipelined NTT paradigm using Montgomery multiplication and a separate pipelined reduction
that is optimized for NTT-friendly primes.
Similarly, PipeZK~\cite{zhang2021pipezk} adopts the pipelined architecture for NTT, utilizing a 4-step NTT pipeline with the goal of dividing larger NTTs to several smaller ones. PipeNTT~\cite{9802731} structurally resembles the pipelined NTT template but introduces a different dataflow that reorders elements after each stage. This reordering aims to reduce the latency caused by buffering in each stage of the original pipelined NTT architecture. 
The design of~\cite{nguyen2024high} uses a multi-path pipelined architecture that provides higher throughput at the cost of increased hardware cost. In this architecture, multiplication and reduction are performed in separate stages: multiplication uses carry-save arithmetic, while reduction leverages the K-RED algorithm~\cite{li2022high}.

\begin{figure}[t]
\centering
\includegraphics[width=0.98\columnwidth]{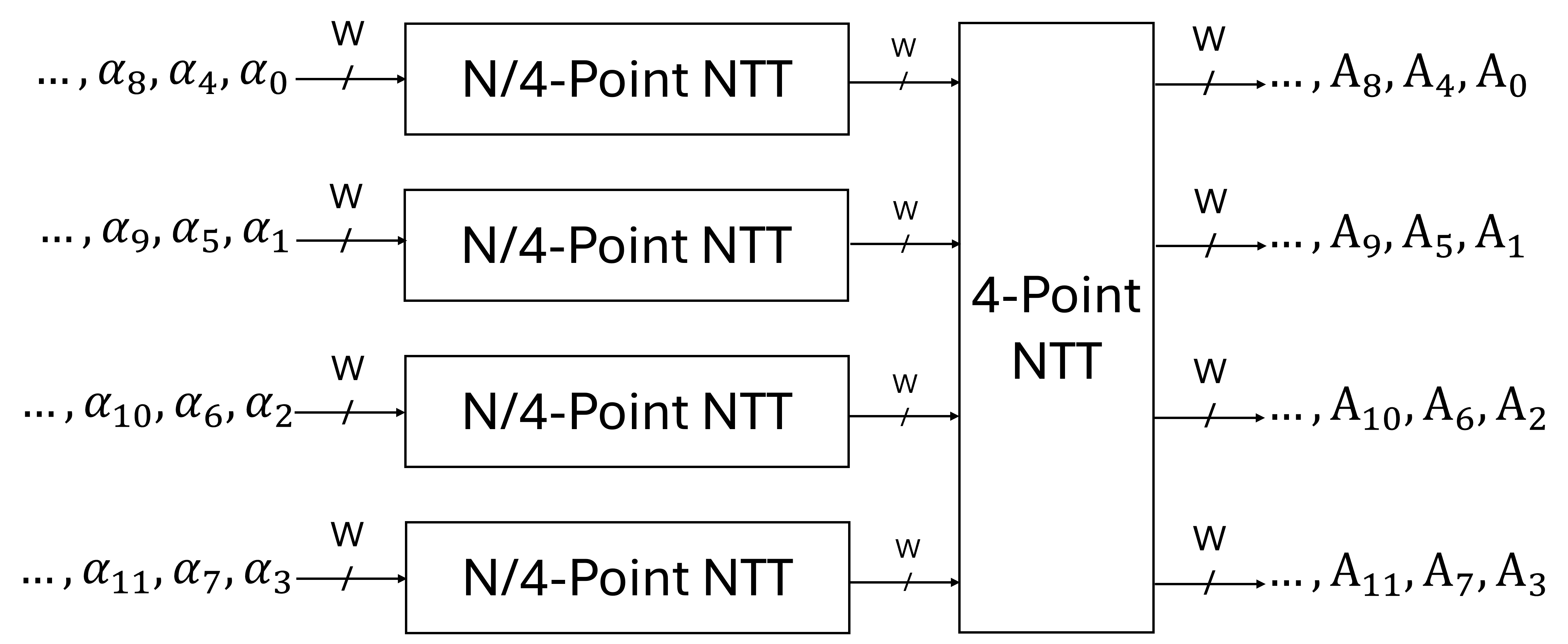}
\caption{Multipath $N$-point pipelined NTT: N/4-point NTT units follow the pipelined design paradigm of Fig.~\ref{fig:sdf}, while the last stage that includes a 4-point NTT follows the fully-parallel butterfly-based design of Fig.~\ref{fig:nttflow}.}
\label{fig:mdf}
\end{figure}

Multipath architectures follow a hybrid approach: Part of the design follows the pipelined paradigm for computing in parallel multiple smaller scale NTT while the outcome of the smaller NTTs is merged in the last stage using a smaller fully-parallel NTT. This design paradigm is highlighted in Fig.~\ref{fig:mdf}.

GS-MDC~\cite{10604827} reduces execution time by leveraging delay of multiplication to reduce the time spent in the buffers. Also, in~\cite{11030819} full utilization is achieved by interleaving two NTTs on the same datapath. This approach is practically useful since two NTTs should be computed for each polynomial multiplication. Difference pipelined dataflow has also been explored. For instance, in~\cite{10710157} NTT stages 
are separated by RAMs, where data are reordered after computation to simplify the memory access of the next stage by pairing the elements that will be used in the same butterfly operation in the next stage.

In all state-of-the-art approaches, the data width throughout the pipeline must adhere to the full width $W$ dictated by the selected modulo $Q$. In \cite{cheon2017homomorphic, parno2016pinocchio} for example, this can result in data sizes of several hundred bits. 
While using narrower data types via digit-serial arithmetic is possible within modulo arithmetic operators, full words must be transferred between arithmetic operators to enable the needed modulo reduction.  
This inconsistency in the size of the data introduces frequent stalls and pipeline bubbles, significantly reducing pipeline efficiency.

\section{The Proposed Architecture for Digit-Serial Pipelined NTTs}
\label{s:proposed}
In this work, our goal is to design a modular, pipelined NTT accelerator which, ---to the best of our knowledge---
leverages digit-serial systolic Montgomery multipliers for the first time. 
To enhance the regularity of the design, we demonstrate how conditional modulo reductions, an inherent part of all modular arithmetic operations in NTT, can be deferred to the final stage of the NTT computation. This transformation enables the design of the entire NTT accelerator using a homogeneous, arbitrarily chosen digit-size. The specific choice of digit-size depends on the required latency, clock frequency, and hardware complexity in terms of area and power consumption.

To allow for a uniform digit-size that is much smaller than the full word-width in all parts of the NTT architecture, we need to encode the input data and all intermediate results in such a way that avoids the need for modulo reduction after each arithmetic operation. To achieve this, the well-known Montgomery modulo reduction strategy defined for modulo multiplication is employed and extended to the add(sub)-multiply structure of each NTT stage. Thus, the derived designs can be applied to build single or multipath pipelined NTTs.

\subsection{Montgomery modulo reduction}
In Montgomery multiplication, the numbers must first be expressed in their so-called Montgomery form, which is essentially a multiplication with a value $R$ (known as the Montgomery radix) that is a power of $2$ chosen to be greater than the modulo $Q$ \cite{montgomery}. For any number $a \pmod{Q}$, its Montgomery form is defined as $\bar{a}=aR \pmod{Q}$. A multiplication of two numbers $\bar{a}$ and $\bar{b}$ in Montgomery form results in $p = \bar{a}\bar{b}=aRbR=abR^2$.  

Montgomery reduction is the process that takes this intermediate result $p$ and calculates $\bar{p}=abR\pmod{Q}$, i.e.,  bringing the result $abR^2$ back into its Montgomery form. Based on~\cite{502403}, this can be achieved with the following formula:
\begin{equation}
\bar{p}=\frac{p+Q[p(-Q^{-1})]_R}{R},
\label{e:redc}
\end{equation}
where $[\cdot]_R$ indicates reduction modulo $R$.
Here division and modular reduction by $R$ are trivial because $R$ is a power of $2$. $-Q^{-1}$ is the negated modular inverse of $Q$ modulo $R$, which is precomputed and used for all the computations.

The output of~\eqref{e:redc} is an integer in the range $[0, 2Q - 1]$. Therefore, when the result exceeds $Q$, an additional subtraction is required to bring it back within the desired input dynamic range of $[0, Q - 1]$.
In~\cite{10.1007/3-540-44499-8_23}, it was shown that this conditional subtraction can be eliminated by allowing inputs to reside in a larger dynamic range. Instead of restricting inputs to the range $[0, Q - 1]$, they can be extended to $[0, 2Q - 1]$. Under this approach, it was demonstrated that the output of~\eqref{e:redc} will also lie within the same extended range, thereby avoiding the need for any correction step. This is because both input and output values remain consistently within the range $[0, 2Q - 1]$.

\subsection{Output range analysis per NTT stage}
We plan to apply the Montgomery modulo reduction shown in~\eqref{e:redc} for modular multiplication within the add(sub)-multiply operations present in each NTT stage, as shown in Fig.~\ref{fig:sdf}. 
Additionally, we will derive an appropriate redundant representation for the input data and all intermediate results that would enable the elimination of conditional subtractions following each modular multiplication, not only within a single NTT stage but across all stages. As a result, the final correction can be deferred to a single subtraction at the end of the computation. This approach allows the design of an NTT accelerator with a uniform data width that is smaller than the full word width $W$.

Specifically, an NTT stage in Fig.~\ref{fig:sdf} computes 
\begin{align}
t_{\text{add}} &= (a_i + a_j)\omega^k \pmod{Q}\label{e:tadd} \\
t_{\text{sub}} &= (a_i - a_j)\omega^k \pmod{Q}\label{e:tsub},
\end{align}
where $a_i$ and $a_j$ are results calculated in the previous NTT stage. The input terms $a_i$, $a_j$ and powers of $\omega$ of any NTT stage are assumed to follow the Montgomery representation. Therefore, the operation of~\eqref{e:tadd} is written as:
\begin{equation}
t_{\text{add}} = (\overline{a_i} + \overline{a_j})\overline{\omega^k} = (a_iR+a_jR)\omega^kR=(a_i+a_j)\omega^kR^2
\label{e:non-reduced}
\end{equation}
A similar derivation can be produced for~\eqref{e:tsub} as well. The result for $t_{\text{add}}$ should be reduced in order to obtain its Montgomery form. 

By applying the Montgomery reduction in \eqref{e:redc} to \eqref{e:non-reduced} we finally obtain that:
\color{black}

\begin{equation}
\overline{t_{\text{add}}} = \frac{t_{\text{add}}+Q[t_{\text{add}}(-Q^{-1})]_R}{R}.
\label{e:addmul}
\end{equation}
We can now calculate the output range of $\overline{t_{\text{add}}}$. We know that $a_i$,$a_j$ and $\omega^k$ are in Montgomery form modulo $Q$. Therefore, the following inequalities hold true:
\begin{equation}
    \overline{a_i} < Q \qquad\quad
    \overline{a_j} < Q \qquad\quad
    \overline{\omega^k} < Q
\label{e:bounds}
\end{equation}
The bounds reported in~\eqref{e:bounds} allow us to calculate the upper bound of $t_\text{add}$ defined in~\eqref{e:non-reduced}, as follows:
\begin{equation}
t_{\text{add}} = (a_iR+a_jR)\omega^kR < (Q+Q)Q=2Q^2    
\label{e:addmulbound}
\end{equation}
By definition $R$ is a power of two larger than $Q$, i.e., $Q < R$. Thus, we can rewrite the bound set in~\eqref{e:addmulbound} as follows: 
\begin{equation}
t_{\text{add}} < 2Q^2 \to t_{\text{add}} < 2QR
\label{e:addmulboundr}
\end{equation}
The term $Q[t_{\text{add}}(-Q^{-1})]_R$ in~\eqref{e:addmul} is always less than $QR$ since the $[t_{\text{add}}(-Q^{-1})]_R$ is by definition a residue modulo $R$. Combining this result with~\eqref{e:addmulboundr} we get for~\eqref{e:addmul} that
\begin{equation}
\overline{t_{\text{add}}} \!=\! \frac{t_{\text{add}}+Q[t_{\text{add}}(-Q^{-1})]_R}{R} \! < \! \frac{2QR+QR}{R} \!=\! 3Q
\label{e:fulbound}
\end{equation}

This result, while correct modulo $Q$, must be further reduced to be less than $Q$, meaning that we may need to subtract $Q$ once or twice from the result. This is normally done by comparing the final result with $Q$. Our goal is to define an appropriate redundant representation and completely avoid the need for such a correction per NTT stage.

\subsection{Removing the need for intermediate modulo reduction}
To achieve this goal we allow our numbers to be in a redundant representation~\cite{Bos_Montgomery_2017}. We specifically restrict them in the range $[0,2Q-1]$ instead of $[0,Q-1]$. Under this redundant representation, each integer $x \pmod{Q}$ can be represented by two different values in this range $x$ and $x+Q$.  
We repeat the presented output range analysis assuming that now the inputs belong to $[0,2Q-1]$. Under this condition, the following inequalities hold true:
\begin{equation}
    \overline{a_i} < 2Q \qquad\quad
    \overline{a_j} < 2Q \qquad\quad
    \overline{\omega^k} < 2Q
\label{e:new-input-bounds}
\end{equation}
The upper bound of $t_\text{add}$ in~\eqref{e:addmulbound}, can be redefined as follows:
\begin{equation}
t_{\text{add}} = (a_iR + a_jR)\omega^kR < (2Q+2Q)2Q=8Q^2 
\label{e:redundantaddmulbound}
\end{equation}
Selecting $R$ differently so that $8Q < R$, allows us to write~\eqref{e:redundantaddmulbound} as follows:
\begin{equation}
t_\text{add} < 8Q^2 < QR
\label{e:redundantaddmulboundr}
\end{equation}
Using the newly derived bound~\eqref{e:redundantaddmulboundr} we can limit the output range of~\eqref{e:addmul} as follows:
\begin{equation}
\overline{t_\text{add}} \!=\! \frac{t_{\text{add}}+Q[t_{\text{add}}(-Q^{-1})]_R}{R} \! < \! \frac{QR+QR}{R} \!=\! 2Q
\end{equation}

This result shows that when the inputs to an NTT stage lie within the dynamic range $[0, 2Q - 1]$, and $R$ is chosen to be greater than $8Q$, the add-multiply output of the NTT stage remains within the same dynamic range as its inputs. This allows the add-multiply operation to be performed consecutively across stages without requiring intermediate comparisons and without concern that the resulting values will grow unbounded as more operations are applied.

To apply this redundant representation uniformly in an NTT stage we need also to take care of the subtract-multiply operation $t_{\text{sub}}$ of the NTT stage defined in~\eqref{e:tsub}. To treat subtract-multiply operation equivalently to the add-multiply operation discussed previously, we need to \emph{adjust on purpose} the output ranges of both add-multiply and subtract-multiply operations. Replacing the bounds defined in~\eqref{e:new-input-bounds} to the difference $a_i - a_j$ found in $t_{\text{sub}}$, we can deduce that
\begin{equation}
-2Q < a_i - a_j < 2Q.
\end{equation}
Similarly, for the sum $a_i + a_j$, we know that $0 < a_i + a_j < 4Q$. To align the output range of both the addition and subtraction operations, we introduce an extra term $2Q$ to the subtraction, resulting in: 
\begin{equation} 
0 < a_i - a_j + 2Q < 4Q 
\label{e:subtract-correction}
\end{equation} 
This adjustment ensures that the output range of $a_i - a_j + 2Q$ matches that of $a_i + a_j$ exactly. Since all operations are performed modulo $Q$, adding the $2Q$ term only introduces redundancy at the output of the subtractor without affecting the final result. With this minor adjustment, the output of the multiplier maintains the same dynamic range, as defined in~\eqref{e:redundantaddmulboundr}, regardless of whether it follows an add-multiply or subtract-multiply operation.

\subsection{Digit serial pipelined NTT with uniform digit sizes}

By eliminating the need for intermediate modulo reductions across all NTT stages, word-level computation can be decomposed into multiple serial steps executed on arbitrarily small digits.
The proposed design operates on a uniform digit size across all arithmetic operators of each NTT stage including adders, subtractors and multipliers. This homogeneity allows us to integrate in NTT systolic high-radix Montgomery multipliers~\cite{4176855} consisting of as many processing elements as the number of digits $W/d$ used to split the initial words.

\begin{figure}[t]
\centering
\includegraphics[width=\columnwidth]{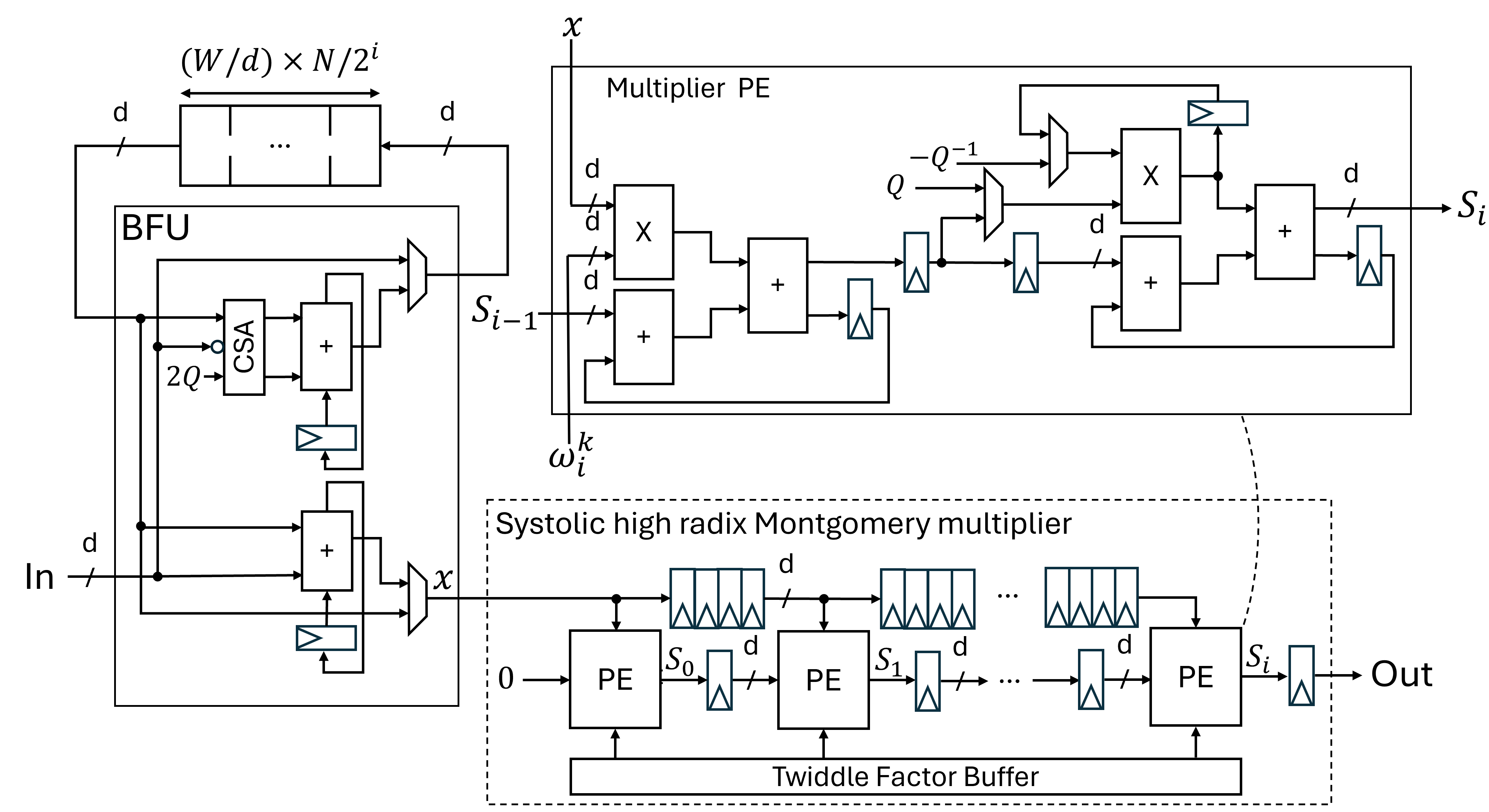}
\caption{The proposed design for the $i$th NTT stage of a single-path pipelined NTT architecture. It utilizes a homogeneous digit-serial architecture, seamlessly integrating a high-radix systolic Montgomery multiplier to achieve modularity and uninterrupted pipelined data flow.}
\label{fig:dsstage}
\end{figure}

Homogeneous digit-serial computation can be performed both to single-path pipelined NTTs that employ a uniform pipelined design similar to Fig.~\ref{fig:sdf} or to multi-path architectures that employ both pipelined and parallel blocks similar to Fig.~\ref{fig:mdf}.  

Fig.~\ref{fig:dsstage} depicts the proposed digit-serial datapath for the $i$th stage of an $N$-point pipelined NTT accelerator. All parts of the design operate on digits with size $d$ bits effectively splitting the initial words into $W/d$ digits.
The adder and subtractor units are replaced by their digit-serial versions. The increment of the subtraction $a_i-a_j$ by $2Q$, as needed by~\eqref{e:subtract-correction}, is implemented using a carry-save adder (CSA) and an adder.
The buffer of the $i$th NTT stage depicted in Fig.~\ref{fig:dsstage}, which originally contained $N/2^i$ words of $W$-bits each, is adjusted to accommodate more $d$-sized digits. Specifically, in this configuration, its size is increased to $(W/d)\times N/2^i $ entries of $d$ bits each.

The Butterfly Unit (BFU) feeds the systolic multiplier with new digit-serial results that pass through $W/d$ digit-serial multiplication PEs following exactly the architecture of~\cite{4176855}. 
Each PE is pipelined to offer single-cycle throughput albeit with a four-cycle latency.
The PE consists mainly of two digit-serial multiply-add units. The first contributes to the multiplication part of the Montgomery algorithm, by calculating a partial product of the two multiplicands and summing it with the previously calculated ones. The second one calculates a partial product of $Q[t_{\text{add}}(-Q^{-1})]_R$ and adds it to the sum. Finally, a division has to be performed according to~\eqref{e:addmul}. Since $R$ is a power of 2, this amounts to a right shift operation and each PE performs a shift by $d$, which in turn results in a total shift by $W$. Since we process one digit of size $d$ per cycle, shifting by $d$ is equivariant to dropping the first digit. This is the reason no extra hardware is needed for the PE to perform this shift, and instead we skip a pipeline register, pushing the sum one cycle forward. 

\begin{figure}[t]
\centering
\includegraphics[width=0.9\columnwidth]{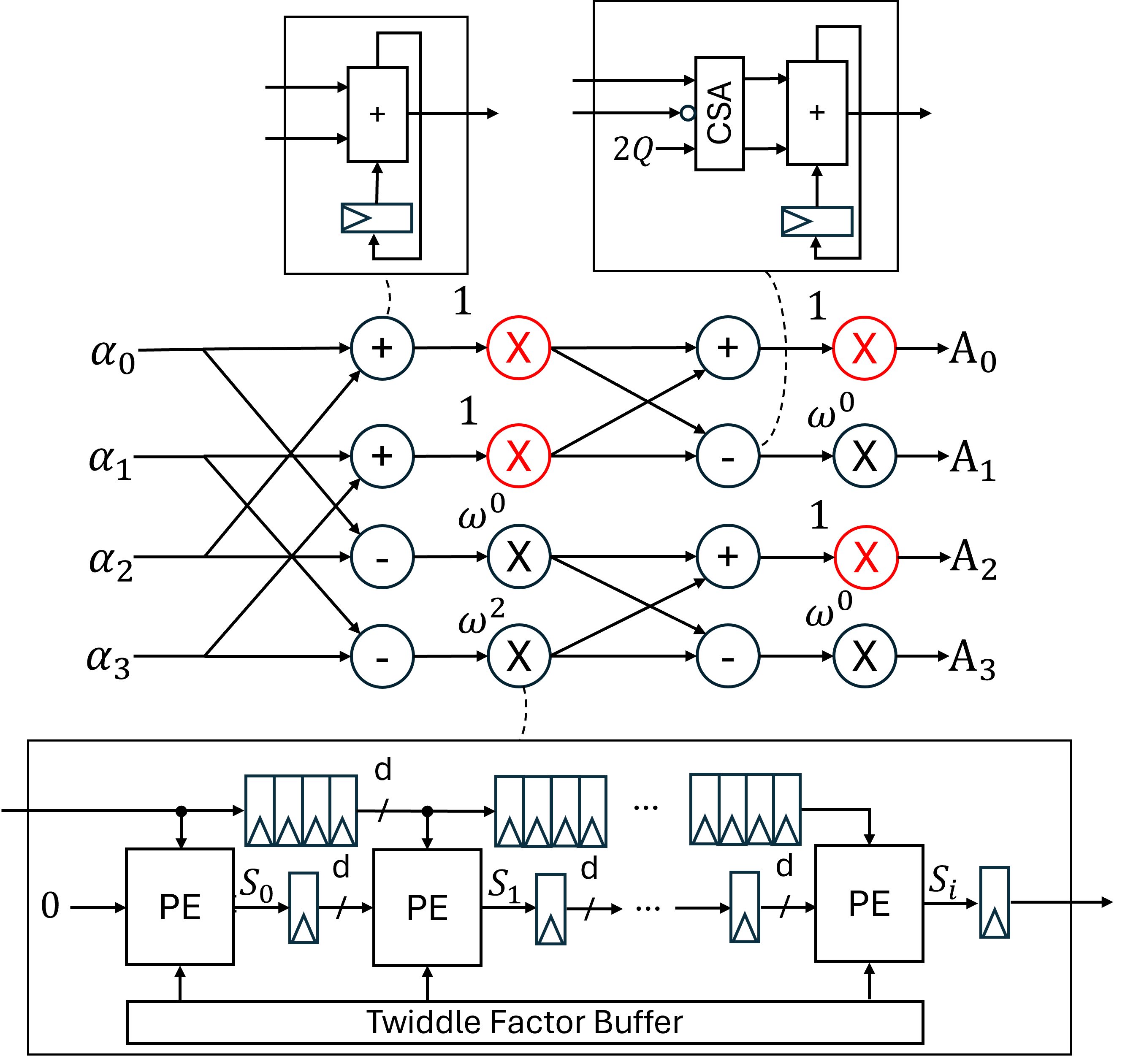}
\caption{A 4-point parallel NTT operating on a digit-serial manner.}
\label{fig:4point}
\end{figure}

The same digit-serial datapath, without the buffer, can be employed in the design of parallel NTT architectures used in the multipath pipelined configurations shown in Fig.~\ref{fig:mdf}. An example of a 4-point parallel NTT utilizing the proposed digit-serial NTTs is illustrated in Fig.~\ref{fig:4point}. To maintain design uniformity, systolic modulo multipliers are included in every branch of the NTT butterfly network. When needed, these multipliers compute the product of the subtraction result and the corresponding twiddle factor. When used solely for uniform modulo reduction, they simply multiply the adder output by 1.

Overall, the design of a multipath pipelined 
$N$-point NTT architecture, which processes words of 
$W$ bits and operates on digits of 
$d$ bits, consists of 
$W/d$ parallel paths. Each path implements an 
$N/(W/d)$-point NTT using a linear pipeline structure. The outputs from all paths are then combined using a 
$W/d$-point parallel NTT at the final stage. Input data elements are distributed across the parallel paths following the same pattern illustrated in Fig.~\ref{fig:mdf}, shown for the 4-path NTT case. The digits of each input element are fed into the NTT serially, one after another.

\section{Experimental results}
\label{s:exp}
This section aims to evaluate the hardware complexity of the proposed digit-serial pipelined NTT in comparison to recent state-of-the-art pipelined NTT architectures, such as Proteus~\cite{hirner2024proteus} and PipeZK~\cite{zhang2021pipezk}. Between those, Proteus employs word-level Montgomery multiplication, which processes an entire word and performs partial reductions over several cycles. This approach aligns more closely with the proposed architecture, making it more suitable for a direct, apples-to-apples comparison. For this reason, we compare our proposed design against an in-house implementation of the Proteus architecture.

\subsection{Comparisons}
According to the Homomorphic Encryption Security Standard~\cite{HomomorphicEncryptionSecurityStandard}, various NTT sizes and word lengths are defined for different security levels. For our comparisons, we evaluate NTT sizes ranging from 128 to 1024 points, using a word length of 
$W=256$ bits. This exceeds the 220-bit requirement specified in the standard, providing sufficient redundancy to satisfy the condition 
$R>8Q$, as required by \eqref{e:redundantaddmulboundr}.

The proposed design supports configurable digit sizes. Since Proteus is assumed to fetch a 256-bit input data element every cycle, we adopt an equivalent input bandwidth for the proposed design by configuring a multipath architecture with 8 parallel paths, each operating on 32-bit digits. Any other configuration satisfying 
$d\times \text{\#paths}=256$ bits would provide the same input bandwidth. Proteus internally performs Montgomery reduction for the 256 width operands in steps of 12-bits. 

\begin{figure}[t]
\centering
\includegraphics[width=\columnwidth]{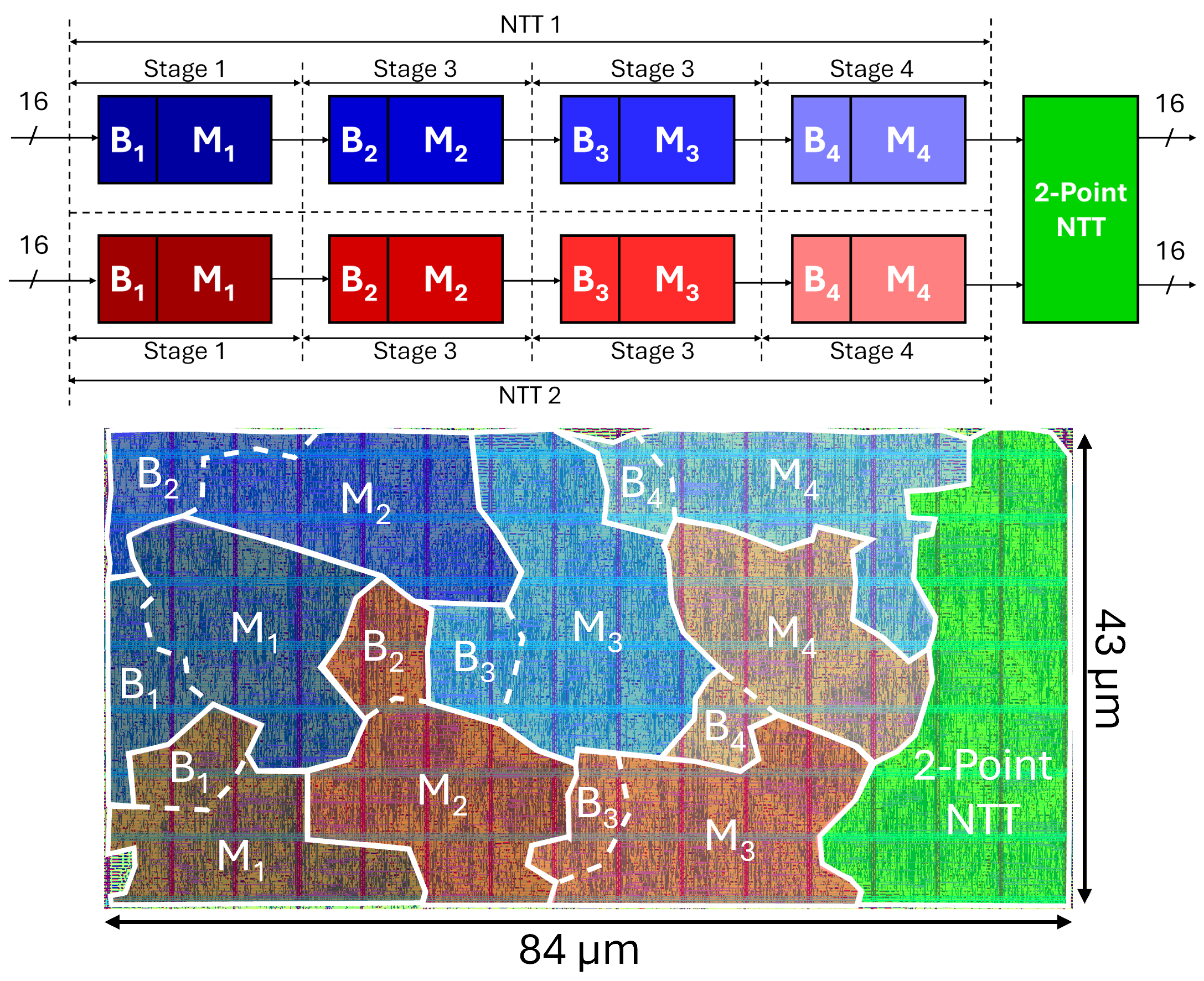}
\caption{The physical layout of a small-scale example of a 32-point NTT following the proposed two-path architecture with  16-bit digits. The first part consists of two pipelined 16-point NTTs with four NTT stages each. The outcomes of the first part are merged at the 2-point NTT of the second part. Each NTT stage consists of a buttefly (B) and a multiplier (M) similar to Fig.~\ref{fig:dsstage}.}
\label{fig:layout}
\end{figure}

Both designs were implemented in SystemVerilog and synthesized down to the physical layout level using the OpenRoad physical synthesis flow~\cite{openroad}, utilizing the ASAP 7 nm standard-cell library. The System Verilog RTL of the proposed design and the utilized flow is open sourced in Github~\footnote{https://github.com/nokia/NTT-Hardware-Accelerator}.  
An example physical layout of small scale NTT that follows the proposed architecture is shown in Fig.~\ref{fig:layout}. This example corresponds to a 32-point NTT implemented using a two-path pipelined architecture. Each path comprises two pipelined 16-point NTT stages operating on 16-bit digits, with the outputs merged by a 2-point parallel NTT. Each stage in the linear pipeline follows the structure shown in Fig.~\ref{fig:dsstage}, consisting of Butterfly Units (B) and Systolic Multipliers (M). The final stage uses the parallel architecture shown in Fig.~\ref{fig:4point}, which uses the same digit-serial adders, subtractors, and multipliers.

The uniform digit size of 32 bits relative to the whole word length multiplication of Proteus, and the smooth integration of the systolic Montgomery modulo multipliers in each NTT stage lead to a significantly increased clock frequency for the proposed design. Specifically, the proposed architecture achieves a clock frequency of 600 MHz, compared to a maximum of 400 MHz for Proteus.

\begin{figure}[t]
    \centering
    \includegraphics[width=0.8\columnwidth]{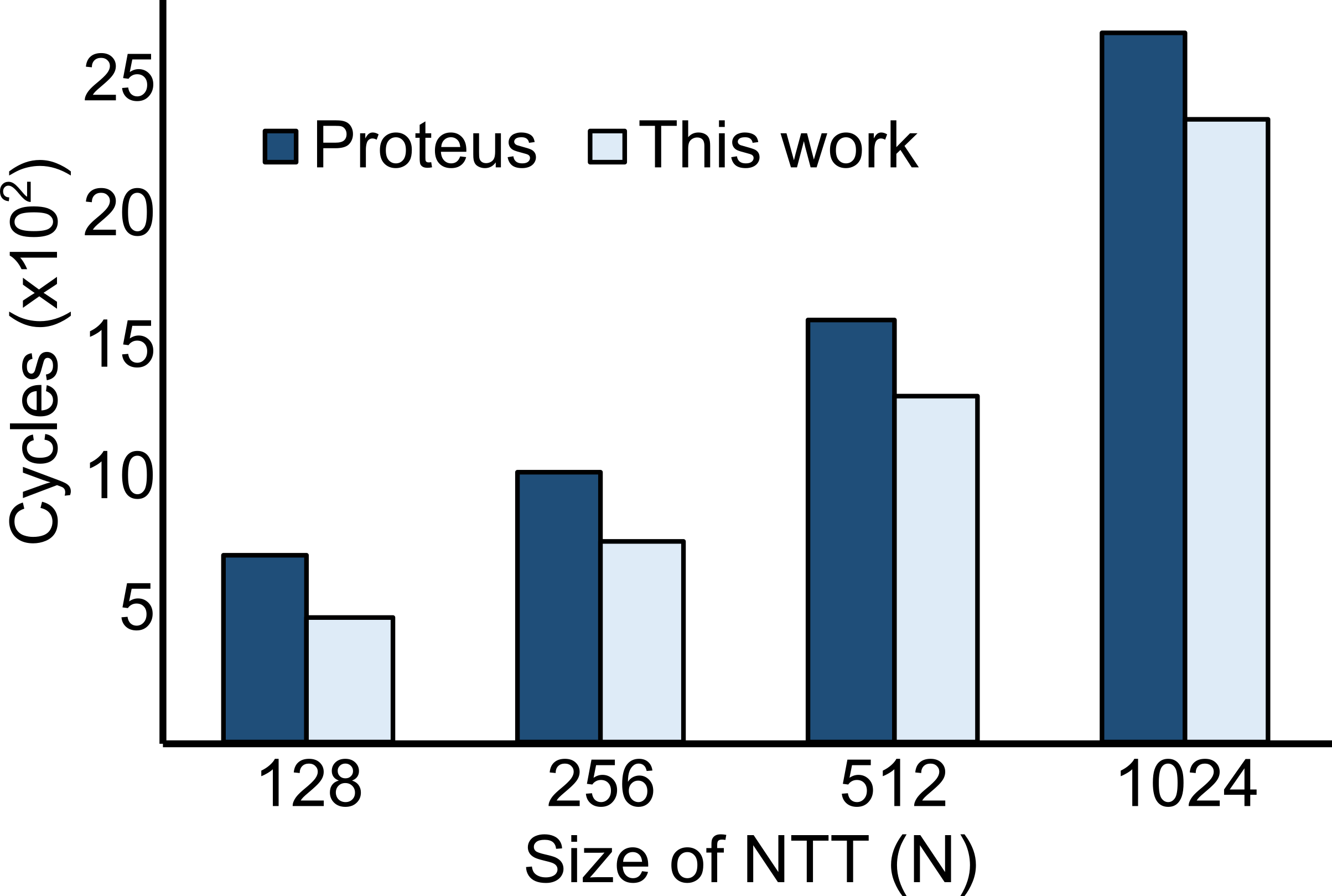}
    \caption{Total execution time in clock cycles for both designs under comparison for various NTT sizes.}
    \label{fig:cycles}
\end{figure}

The multipath architecture of the proposed design allows it to complete NTT computation in less cycles than a purely pipelined one. Figure~\ref{fig:cycles} shows the total number of cycles each design requires to perform the NTT for various sizes of  $N$. The proposed design consistently completes the computation faster across all values of $N$. When combined with its higher operating frequency, this results in a substantial reduction in overall latency.

\begin{figure}[t]
\centering
\includegraphics[width=0.8\columnwidth]{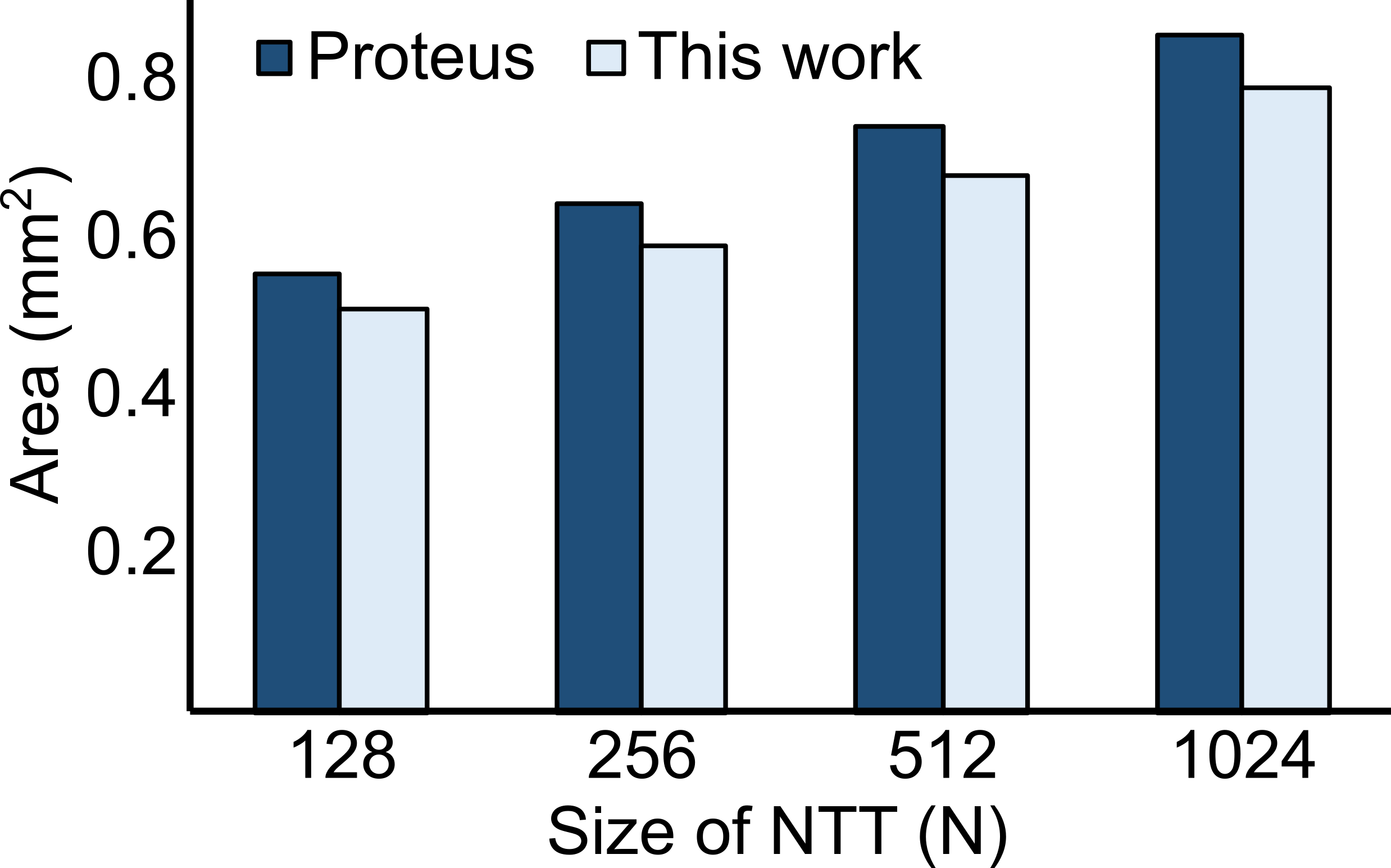}\\
{(a)}\\
\includegraphics[width=0.76\columnwidth]{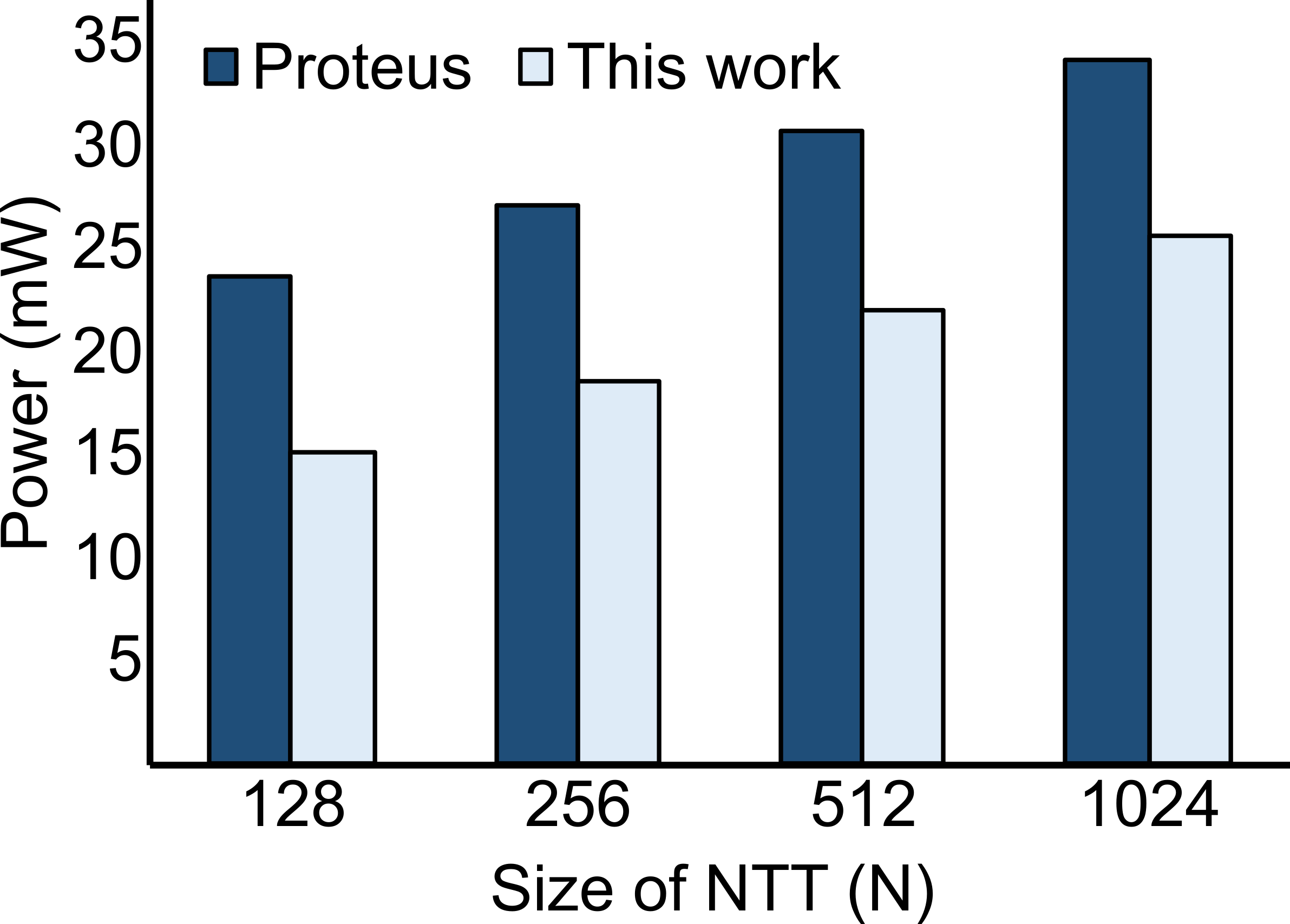}    \\
{(b)}
\caption{The (a) area and (b) power of both designs under comparison for various NTT sizes. }
\label{fig:area-power}
\end{figure}

In addition to being faster, the proposed design also requires less hardware area than Proteus. Figure~\ref{fig:area-power}(a) presents a comparison of the area requirements for both designs across various NTT sizes. The proposed design achieves area savings ranging from 7.9\% to 8.7\%.

A similar trend is observed in terms of power consumption, as shown in Fig.~\ref{fig:area-power}(b), where the proposed design achieves power savings ranging from 25\% to 36\%. The most significant reduction in power consumption stems from the use of a systolic Montgomery multiplier that operates with a uniform digit size and eliminates additional reduction steps, thanks to the chosen redundant data representation. The remaining parts of the design, specifically, the buffers and pipeline registers common to any pipelined NTT, remain identical in both architectures. Due to clock gating, the power overhead of the buffers is minimal. As a result, reducing the power cost associated with multiplication, which involves complex combinational logic, has a more substantial impact on the overall efficiency of these designs.

\subsection{Scaling of Hardware Complexity with Digit Size}
Finally, we examine how the hardware complexity of the proposed design scales with different digit sizes. Reducing the digit size simplifies the combinational logic in each pipeline stage, which is expected to result in higher clock frequencies. The total execution time in cycles remains nearly constant, as the increased number of parallel paths in the pipelined NTT offsets the latency introduced by greater serialization. However, this additional parallelism incurs a trade-off in the form of increased area.

\begin{figure}[t]
\centering
\includegraphics[width=0.8\columnwidth]{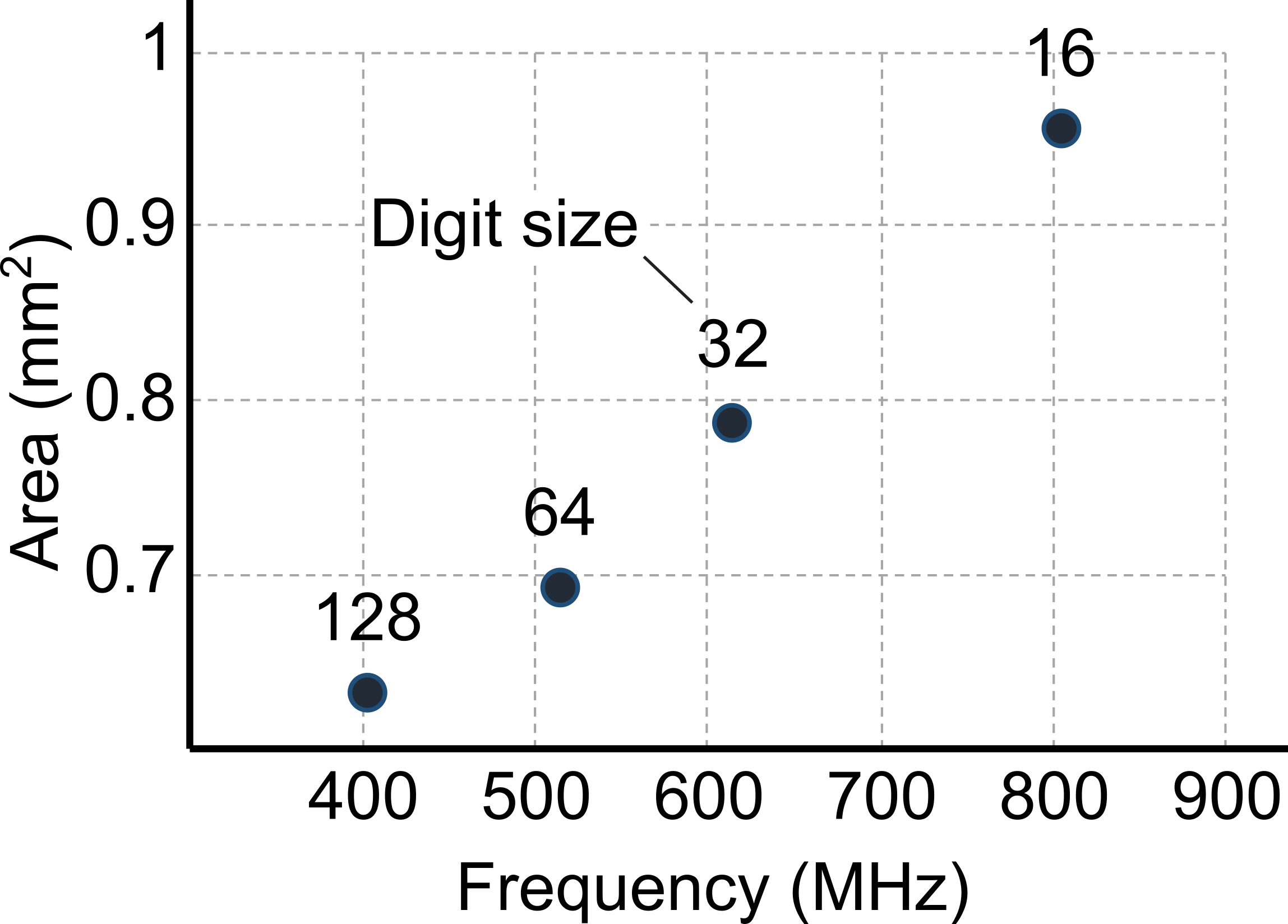} \\
{(a)}\\
\includegraphics[width=0.8\columnwidth]{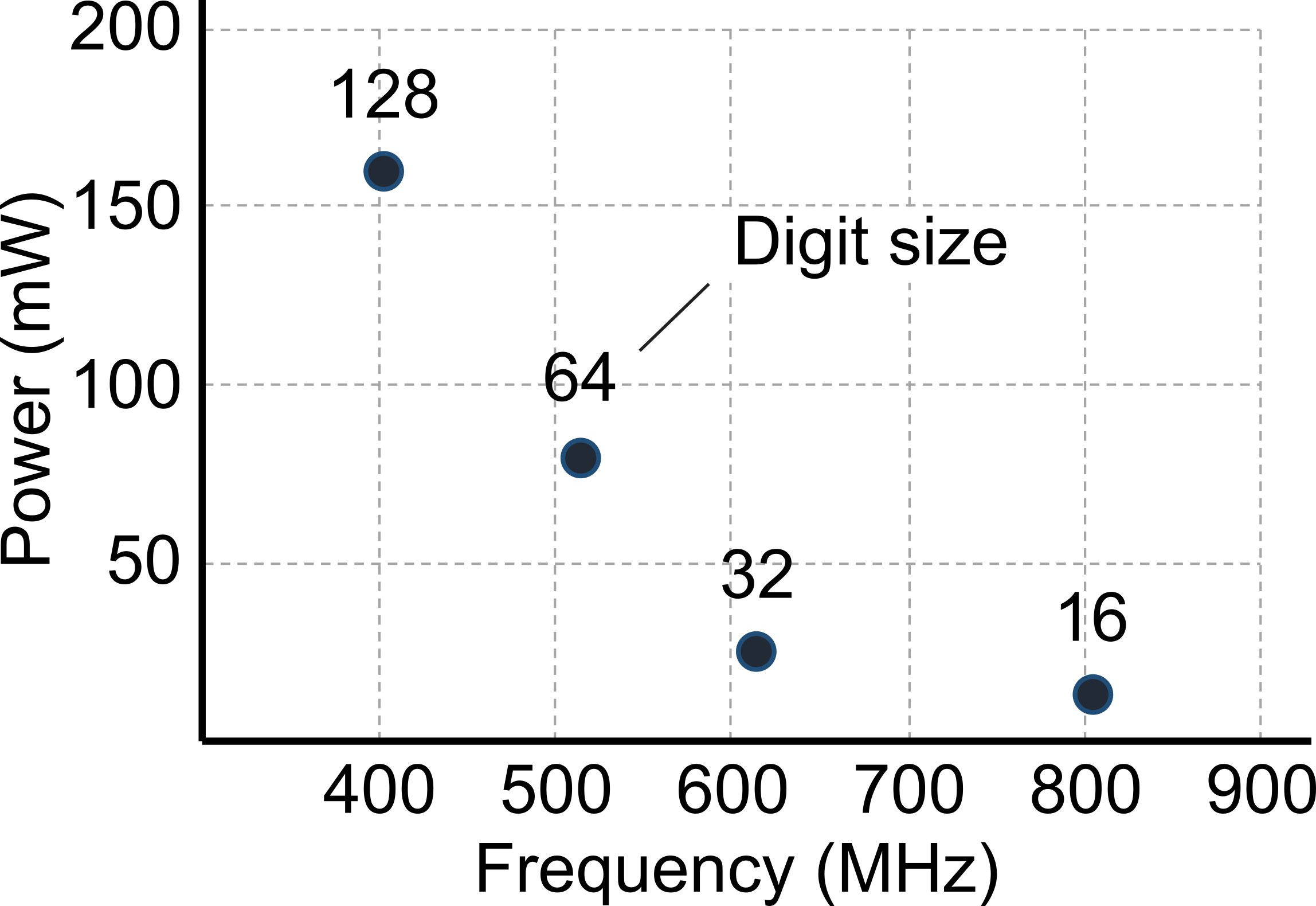} \\
{(b)}
\caption{The (a) area, (b) power and maximum achievable clock frequency for a 1024-NTT using the proposed architecture for various digit sizes.}
\label{fig:area-freq}
\end{figure}

These assumptions are confirmed by the hardware complexity results shown in Fig.~\ref{fig:area-freq} for a 1024-point NTT. The area scales nearly linearly with clock frequency across different digit sizes. All configurations maintain the same input-output bandwidth constraint of 256 bits per cycle. This flexibility allows designers to select a digit size that best balances clock frequency and area requirements. For example, reducing the digit size from 128 bits to 16 bits nearly doubles the maximum clock frequency while increasing the area of the NTT accelerator by approximately 33\%.

Power consumption behaves differently across digit sizes. While area increases with smaller digit sizes, power consumption actually decreases. This is because increasing the digit size raises the power consumption of each processing element (PE) in the multiplier more rapidly than it reduces the total number of PEs required per multiplier. Additionally, although buffers significantly impact area, their contribution to dynamic power is minimal due to clock gating, only one buffer is accessed (read or written) per cycle in each NTT stage, while the others remain gated. This effect is especially pronounced in large NTT sizes, such as our 1024-point design, which includes 64K flip-flops used for buffering. As a result, larger digit sizes ultimately lead to increased overall power consumption.

\section{Conclusions}

This work presents a scalable approach to NTT acceleration based on digit-serial arithmetic and uniform digit sizes. By leveraging existing efficient modular arithmetic units and a carefully selected redundant data representation, we design modular pipelined NTT architectures that avoid serialization overhead and intermediate modular reductions throughout the pipeline. The integration of systolic digit-serial Montgomery multipliers operating under a uniform digit size enables high-speed, power-efficient implementations that outperform recent state-of-the-art both in terms of achievable clock frequency as well as power consumption under the same execution time.

\section*{Acknowledgments}
This work is supported by a research grant of Nokia Bell Labs to Democritus University of Thrace for "Hardware accelerators for Privacy Preserving Technologies"

\bibliographystyle{IEEEtran}
\bibliography{main}
\end{document}